\documentclass[conference]{IEEEtran}
\IEEEoverridecommandlockouts

\usepackage[T1]{fontenc}
\usepackage{graphicx}
\usepackage{tikz}
\usepackage{amsmath}
\usepackage{url}
\usepackage{textcomp}
\usepackage{balance}
\usepackage[most]{tcolorbox}
\usepackage{enumitem}
\usepackage{caption}
\usepackage{threeparttable}
\usepackage{multirow}
\usepackage{makecell}
\usepackage{cite}
\usepackage{hyperref}
\usepackage{array}
\usepackage{xcolor}
\usepackage{makecell}
\usepackage[caption=false,font=footnotesize]{subfig}
\usepackage{booktabs}
\usepackage{amssymb}   
\usepackage{pifont}    

\newcommand{\sstitle}[1]{\noindent{\underline{\textit {#1}}}}

\newcommand{\add}[1]{\textcolor{black}{\textbf{} #1}}

\begin{document}
%
\title{ESG Reporting Lifecycle Management with Large Language Models and AI Agents}

\author{
\IEEEauthorblockN{
Thong Hoang\IEEEauthorrefmark{1}\IEEEauthorrefmark{4},
Mykhailo Klymenko\IEEEauthorrefmark{1}\IEEEauthorrefmark{4},
Xiwei Xu\IEEEauthorrefmark{1},
Shidong Pan\IEEEauthorrefmark{1}, \\
Yi Ding\IEEEauthorrefmark{2},
Xushuo Tang\IEEEauthorrefmark{2},
Zhengyi Yang\IEEEauthorrefmark{2},
Jieke Shi\IEEEauthorrefmark{3},
David Lo\IEEEauthorrefmark{3}
}

\IEEEauthorblockA{
\IEEEauthorrefmark{1}CSIRO's Data61, Australia; 
\IEEEauthorrefmark{2}University of New South Wales, Australia;
\IEEEauthorrefmark{3}Singapore Management University, Singapore
}
\IEEEauthorblockA{\IEEEauthorrefmark{4}
Authors contributed equally to this work.
}
}

%
%
%
%
%
\maketitle              
\begin{abstract}

Environmental, Social, and Governance (ESG) standards have been increasingly adopted by organizations to demonstrate accountability towards ethical, social, and sustainability goals. However, generating ESG reports that align with these standards remains challenging due to unstructured data formats, inconsistent terminology, and complex requirements. Existing ESG lifecycles provide guidance for structuring ESG reports but lack the automation, adaptability, and continuous feedback mechanisms needed to address these challenges. 
To bridge this gap, we introduce an agentic ESG lifecycle framework that systematically integrates the ESG stages of identification, measurement, reporting, engagement, and improvement. In this framework, multiple AI agents extract ESG information, verify ESG performance, and update ESG reports based on organisational outcomes. By embedding agentic components within the ESG lifecycle, the proposed framework transforms ESG from a static reporting process into a dynamic, accountable, and adaptive system for sustainability governance. We further define the technical requirements and quality attributes needed to support four main ESG tasks, such as report validation, multi-report comparison, report generation, and knowledge-base maintenance, and propose three architectural approaches, namely single-model, single-agent, and multi-agent, for addressing these tasks. The source code and data for the prototype of these approaches are available at \url{https://gitlab.com/for_peer_review-group/esg_assistant}.


\end{abstract}

\section{Introduction}
\label{sec:intro}

\textbf{ESG} stands for Environmental, Social, and Governance, which are dimensions to evaluate organisations' sustainability aspects and ethical impacts in our society. The environmental dimension assesses how organizations mitigate climate change effects and control pollution. The social dimension examines organizational social standards related to diversity and inclusion as well as human rights. The governance dimension evaluates an organization’s approach to adherence to regulations and ethical decisions. Organizations have started to focus on ESG after realizing that it supports their long-term success~\cite{glowik2024blackrock}. 
In 2025, ESG disclosures become mandatory for organisations in many countries (e.g., the U.S., the United Kingdom, and Singapore), requiring them to report their sustainability impacts to evaluate the organizations' sustainability performance and risks.\footnote{\url{https://www.presgo.com/resources/global-mandatory-esg-disclosures/}} 

Existing ESG standards~\cite{GRI, SASB} assist organisations in estimating their ESG performance and maintaining accountability, reliability, and transparency in their ESG reporting. 
However, there are several challenges in automating, verifying, and aligning ESG reports with the ESG standards across organisations. 

\sstitle{Unstructured and heterogeneous formats.} ESG data exists in varied formats, e.g., tables, narratives, scanned documents, and multimedia, making consistent extraction difficult. Optical Character Recognition (OCR) and rule-based systems often introduce errors, reducing reliability.

\sstitle{Inconsistent terminology and irrelevant content.}  Differences in terminology (e.g., ``greenhouse gas emissions'' vs. ``carbon footprint'') and reporting granularity impede normalization. Moreover, ESG reports blend substantive information with promotional content, complicating automated filtering.

\sstitle{Multi-standard alignment.}  Mapping ESG reports across multiple ESG standards remains non-trivial due to divergent scopes and taxonomies.


\sstitle{Updating ESG standards and regulatory pressure.}  Frequent updates to global ESG standards, i.e., SASB~\cite{SASB}, require adaptive models capable of incorporating new taxonomies and compliance requirements.

Various ESG lifecycles~\cite{InvestEuropeESGLifecycle, esrs_lifecycle_2025} have been proposed to help organisations align their ESG information with ESG standards; however, they do not address the challenges in generating ESG reports. These challenges, such as unstructured data formats, inconsistent terminology, and complex multi-standard alignment, require an automated and intelligent framework to process and interpret ESG data. Existing lifecycles are domain-specific, lack mechanisms for continuous feedback, and depend on manual data collection and interpretation. As a result, they are unable to address the challenges inherent in ESG reporting.

To address these issues, we present an agentic ESG lifecycle framework designed to enhance automation, explainability, and adaptability in assisting organizations to follow ESG standards. The proposed framework includes five stages: identification, measurement, reporting, and engagement. To operationalize this lifecycle, we embed Large Language Models (LLMs) as agentic components across all stages. These agentic components enable automation and explainability by extracting ESG data from various sources and aligning them with ESG standards to automatically generate ESG reports and update them based on sustainability performance.


Based on the ESG lifecycle, we present technical requirements and quality attributes necessary to operationalize an LLM-based ESG framework. Specifically, we define four primary tasks: (1)~report validation and compliance checking, (2)~multi-report comparison and trend evaluation, (3)~automated report generation, and (4)~maintenance of a knowledge base of ESG standards and regulations. Each task can be employed by LLMs to improve their accuracy and efficiency. Apart from traditional quality attributes, i.e., modularity, flexibility, scalability, autonomy, and robustness, we also identify LLM quality attributes, such as explainability, context relevance, and answer faithfulness, to improve transparency and trustworthiness of the LLM-based ESG framework. 


Our paper also introduces three architectural approaches for implementing the LLM-based ESG framework in these tasks. First, a single-model architecture is designed where one LLM handles all tasks by prompting. Second, a single-agent architecture approach employs retrieval-augmented generation (RAG) and external tools, including data parsers, ESG databases, and sentiment analyzers, to present results. Finally, a multi-agent architecture integrates a specific agent for each ESG task, such as validation, comparison, and reporting. In the proposed single-agent and multi-agent architectures, RAG is a core knowledge base component to ensure accuracy and explainability for ESG tasks.

In summary, the main contributions of our paper are as follows:

\begin{itemize}[leftmargin=*]
    \item Proposes an agentic ESG lifecycle framework that integrates LLMs across all lifecycle stages to enable automation, adaptability, and explainability in the ESG platform.
    \item Identifies four main ESG tasks: (1) report validation, (2) report comparison, (3) report generation, and (4) knowledge base maintenance. We then define the technical requirements and quality attributes necessary for implementing an LLM-based ESG framework.
    \item Proposes three different architectural approaches for implementing these four main tasks.
\end{itemize}
\section{Proposed Framework and ESG Discovery}
\label{sec:method}

\begin{figure}[t!]
    \centering
    \includegraphics[width=0.475\textwidth]{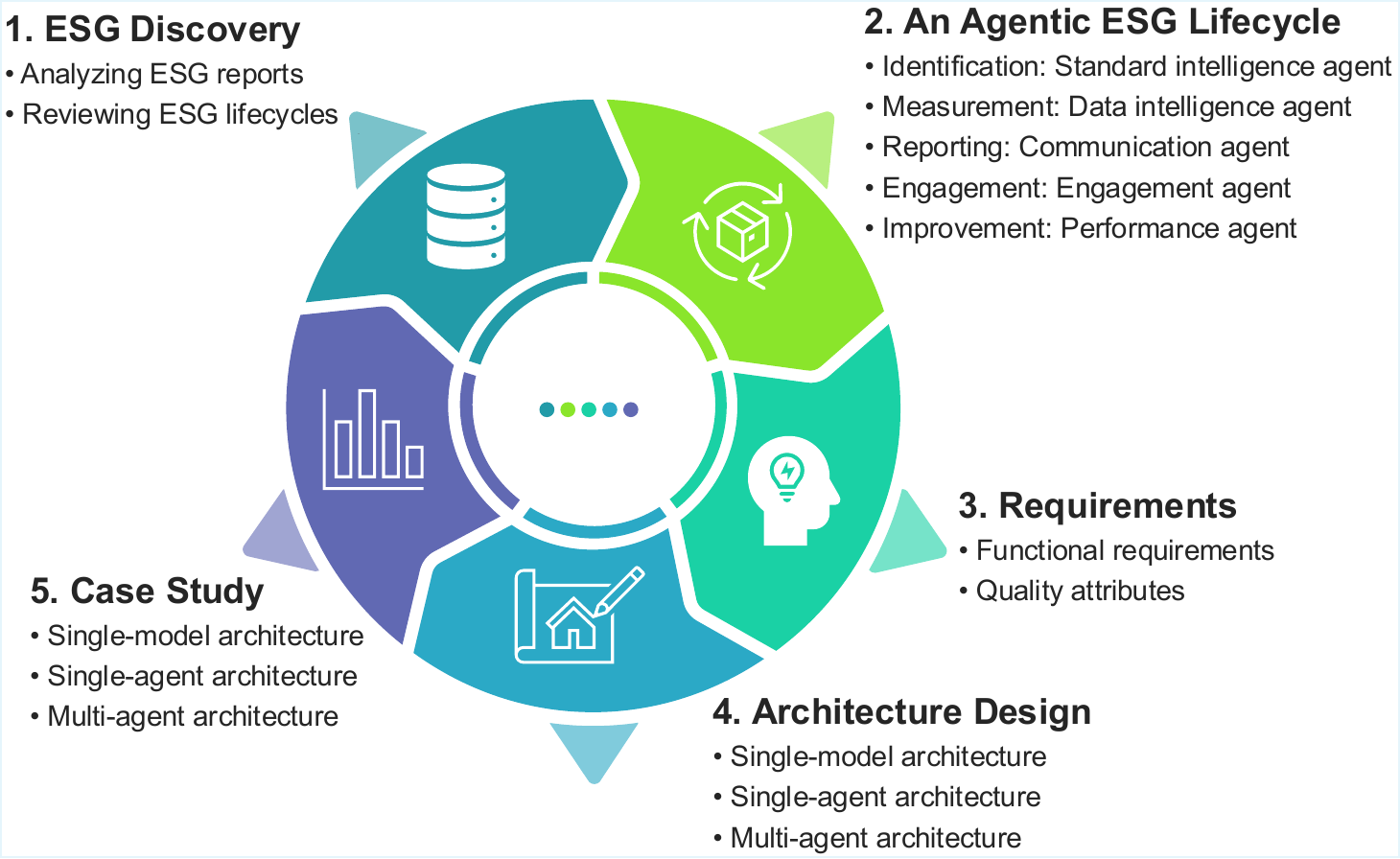}
    \caption{Overview of the proposed framework in our research paper.}
    \label{fig:method}
\end{figure}

\subsection{Proposed Framework}

Existing research on sustainability and ESG reporting~\cite{sroufe2017integration, rezaee2017business} identifies several components for enabling organizations to produce transparent and reliable ESG reports for investment decision-making. Figure~\ref{fig:method} presents an overview of our proposed framework. Specifically, it includes five core components: 

\begin{enumerate}[leftmargin=*]
    \item \textit{ESG discovery:} Our methodology begins by analyzing ESG reports from organisations to understand their ESG reporting structures and disclosed ESG metrics. Moreover, we also review existing ESG lifecycles to understand their limitations.
    \item \textit{ESG lifecycle refinement:} We introduce an agentic ESG lifecycle framework designed to enhance automation, explainability, and adaptability in organisational adherence to ESG standards. It also facilitates the decision-making process for investors and stakeholders. 
    \item \textit{Requirements:} Based on this agentic lifecycle, we define the functional requirements and quality attributes, i.e., modularity, flexibility, and usability, necessary to support ESG tasks, such as report validation, multi-report comparison, and report generation. 
    

    \item \textit{Architecture design:} We propose three architectural approaches for implementing an LLM-based framework to address ESG tasks. First, we introduce a single-model architecture in which a single LLM is responsible for all tasks by relying on prompt engineering. Second, we present a single-agent architecture that integrates retrieval-augmented generation and external tools, i.e., data parsers or sentiment analyzers, to solve these tasks. Lastly, we propose a multi-agent architecture in which we define a specific agent for each ESG task. 
    
    \item \textit{Case study:} We implement a prototype for each of the proposed architectures, such as the single-model, single-agent, and multi-agent architectures, focusing on the ESG report validation and compliance checking task. In this case study, human manual verification serves as the baseline. We compare the effectiveness of the three architectures using three evaluation metrics, i.e., accuracy, computational cost, and energy consumption.
\end{enumerate}

\subsection{ESG Discovery}
\label{sec:data}

The ESG discovery is the first component of our research methodology. Its goal is to analyze ESG reports to understand how organizations align their ESG reports with ESG standards, i.e., GRI or SASB. This component also includes a review of existing ESG lifecycles, which helps identify their limitations and motivates the development of our agentic ESG lifecycle.

\begin{table*}[t!]
\centering
\caption{Summary of selected ESG reports. The Global Reporting Initiative~\cite{GRI}, Sustainability Accounting Standards Board~\cite{SASB}, and Task Force on Climate-Related Financial Disclosures~\cite{TCFD} standards are represented as [G], [S], and [T], respectively. Combined labels indicate overlap across standards, i.e., [GS] or [GST]. Symbols indicate reporting challenges: † multi-standard alignment, ‡ heterogeneous or unstructured disclosure formats.}
\label{tab:sustainability-summary}
\small
\begin{tabular}{
p{0.08\linewidth}p{0.27\linewidth}p{0.27\linewidth}p{0.27\linewidth}}
\toprule
\textbf{Organisation} & \textbf{Environmental Dimension} & \textbf{Social Dimension} & \textbf{Governance Dimension} \\
\midrule

BP†‡ &
- Scope 1+2: $\downarrow$35\% vs.\ 2019 [GST] \newline
- Carbon investment: \$2.2B [S] \newline
- Renewables pipeline: 23\,GW [GST] \newline
- Hydrocarbon flaring: $\uparrow$967kt [GS]
&
- Women in leadership: 31\% [GS] \newline
- Vehicle accidents: $\downarrow$20\% [G] \newline
- Oil spills: 121 [GS] \newline
- Concerns raised: $\downarrow$14\% [G]
&
- Sustainability committee meetings [G] \newline
- Human rights policy [GS]
\\
\hline

Microsoft†‡ &
- Scope 1+2: $\downarrow$17\% [GST] \newline
- Carbon removal: 1,443,981 tons [G] \newline
- Waste diversion: 12,159 tons [G] \newline
- Clean water: 1,000,000 people [G]
&
- Supplier decarbonization [G] \newline
- Clean energy access [G] \newline
- Land protection [G]
&
- Sustainability committee review [T] \newline
- Internal carbon fee applied [T] \newline
- Sustainability scorecards [G]
\\
\hline

Dell†‡ &
- Scope 1+2 $\downarrow$44\% [GST] \newline
- Renewable electricity: 61.5\% [GS] \newline
- Recycled materials: 14.1\% [G] \newline
- E-waste recovered: 90,000 tons [S]
&
- Women in leadership: 29.1\% [GS] \newline
- Inclusive culture: 16.1\% [GS] \newline
- Health \& well-being [G] \newline
- Work-related fatalities: 0 [G]
&
- Customer security audits [GS] \newline
- ESG steering committee [GS] \newline
- Human rights committee [GS]
\\
\hline

Pfizer†‡ &
- Scope 1+2: $\downarrow$13.9\% [GST] \newline
- Energy consumption [GST] \newline
- Climate performance KPI [T] \newline
- Climate risk analysis [T]
&
- Patients treated: 525.2M [G] \newline
- Women at VP+: 44.8\% [GS] \newline
- Product donation programs [GS] \newline
- Health workers/facilities [G]
&
- Compensation [G+S] \newline
- Regulatory \& compliance committee [G] \newline
- ESG progress reporting [G+S]
\\
\hline

NVIDIA†‡ &
- Renewable energy target [GST] \newline
- GPU efficiency: 20$\times$ CPU [S] \newline
- Scope 1+2+3 disclosed [GST] \newline
- Landfill diversion: 71\% [G]
&
- Audit coverage: 50\% [GS] \newline
- Strategic suppliers: 100\% [GS] \newline
- Code of conduct compliance [GS] \newline
- Parental leave: 22 weeks [G]
&
- TCFD framework implemented [T] \newline
- Sustainability committee [GS] \newline
- Annual supplier report [GS]
\\
\hline

\end{tabular}
\end{table*}

\noindent \textit{1. Analyzing ESG reports:} 
\add{We manually reviewed 13 ESG reports\footnote{\url{https://gitlab.com/for_peer_review-group/esg_assistant}} and conducted a structured content analysis to examine their reporting structures and disclosed metrics. For each report, we extracted its ESG factors and mapped them to GRI, SASB, and TCFD standards. We categorized these factors into three ESG dimensions, i.e., environmental, social, and governance.
Table~\ref{tab:sustainability-summary} presents selected ESG reports from major global organisations, illustrating how their ESG metrics align with ESG standards, namely GRI~\cite{GRI}, SASB~\cite{SASB}, and TCFD~\cite{TCFD}. Each selected organization represents an individual domain, e.g., energy for BP;\footnote{\url{https://www.bp.com/}} software for Microsoft;\footnote{\url{https://www.microsoft.com/}} hardware for Dell\footnote{\url{https://www.dell.com/}} and NVIDIA;\footnote{\url{https://www.nvidia.com/}} biotechnology for Pfizer.\footnote{\url{https://www.pfizer.com.au/}} We select these reports due to their maturity and transparency in disclosing ESG information. As shown in Table~\ref{tab:sustainability-summary}, the comparison of ESG metrics across 13 ESG reports reveals two primary challenges that organisations face when aligning their ESG data with multiple standards as follows.}

\begin{table*}[t!]
\centering
\caption{Summary of ESG lifecycles}
\label{tab:esg-lifecycle-models}
\begin{tabular}{p{0.18\linewidth}p{0.18\linewidth}p{0.25\linewidth}p{0.29\linewidth}}
\hline
\textbf{Framework} & \textbf{Domain} & \textbf{Stage} & \textbf{Description} \\
\hline

Invest Europe ESG lifecycle tool~\cite{InvestEuropeESGLifecycle} &
Private equity and venture capital investment. &
- Fundraising 

- Due diligence and ownership 

- Monitoring and reporting 

- Exit and distribution
&
Maps ESG dimensions to each stage of the full investment cycle. This framework focuses on alignment between investors and organizations on long-term value creation. \\

\hline
ESRS reporting lifecycle model~\cite{esrs_lifecycle_2025} &
Corporate sustainability reporting under EU standards. &
- Preparation and scoping

- Materiality assessment

- Data collection and control

- Disclosure and assurance
&
Provide a detailed 189-step reporting process for organizations. The framework focuses on regulatory compliance and audit readiness, mainly applying to EU organizations. \\

\hline
Generic ESG-LCSA integration framework~\cite{padilla2025enhancing} &
Academic sustainability science and product life-cycle assessment. &
- Goal and scope definition

- Inventory and impact assessment

- Interpretation and decision support
&
Integrates environmental, social, and economic dimensions into the ESG decision-making process.  
Supports product-level sustainability investment decisions. \\

\hline
ESG maturity model~\cite{PwCESG} &
Organizational ESG governance and capability development. &
- Minimalist

- Pragmatist

- Strategist

- Trailblazer
&
Provides ESG program assessment and roadmap design. 
Aligns ESG data with organizations' values to build the trust of stakeholders. \\
\hline
\end{tabular}
\end{table*}

\sstitle{Multi-standard alignment.} Many ESG metrics appear across multiple standards; however, each standard requires different expectations for the same metric. For example, Scope 1, Scope 2, and Scope 3 greenhouse gas emissions metrics, presented in teal in Table~\ref{tab:sustainability-summary}, reflect the requirements of three standards: GRI~\cite{GRI}, SASB~\cite{SASB}, and TCFD~\cite{TCFD}. However, each standard interprets these metrics differently. SASB requires the quantitative emissions numbers. GRI expects an explanation of how these emissions relate to an organisation's operations. TCFD requires organizations to embed these metrics within a financial climate risk context. As a result, organisations must reinterpret the same metrics multiple times to satisfy diverging expectations from multiple standards.

\sstitle{Unstructured and heterogeneous formats.} The ESG reports in Table~\ref{tab:sustainability-summary} show that ESG metrics are presented in varied formats, requiring substantial manual effort to extract and understand from stakeholders for investment decisions. For example, BP and NVIDIA employ multi-column tables to report key environmental metrics such as greenhouse gas emissions, landfill diversion, and energy efficiency. Microsoft focuses on providing narrative explanations for these metrics, while Dell presents renewable electricity and recycling metrics using infographics and sidebars. Pfizer summarizes Scope 1–3 emissions using figures and charts. The combination of tables, figures, and explanations creates significant challenges for stakeholders, making it difficult to interpret and compare ESG reports across organisations.

\noindent \textit{2. ESG lifecycles:} Table~\ref{tab:esg-lifecycle-models} presents four ESG lifecycle frameworks, each corresponding to a specific domain. 
We identify three limitations of these existing ESG lifecycles. 
\begin{itemize}[leftmargin=*]
    \item Each lifecycle focuses on a specific domain, lacking the flexibility to extend to other domains or organizations. For example, the Invest Europe ESG lifecycle applies only to private equity, while ESRS is employed for EU organizations. 
    \item The ESG lifecycles lack a feedback system that supports organizations in improving the alignment of their ESG information with the lifecycle's outcomes. For instance, the ESG–LCSA framework ends at the decision stage and offers no guidance on how organizations should incorporate the results of those decisions back into earlier stages, limiting opportunities for refinement of aligning the ESG information.
    \item The ESG lifecycles lack automation capabilities for integrating ESG data across their stages. For example, both the ESG maturity model and the ESRS reporting lifecycle outline detailed stages and procedural steps; however, these steps depend on human effort to collect, extract, and align ESG information at each stage.
\end{itemize}

These limitations highlight the need for an agentic ESG lifecycle that can be adopted across organizational domains. The agentic lifecycle integrates continuous feedback mechanisms to improve alignment between ESG data and the various stages of the process based on the decision outcomes. It also includes automation capabilities to collect and extract ESG information across departments and align it at each stage of the lifecycle, thereby reducing manual effort and enhancing ESG reporting efficiency.

\section{An Agentic ESG Lifecycle}
\label{sec:lifecycle}


\begin{figure}[t!]
  \centering
  \includegraphics[width=0.925\linewidth]{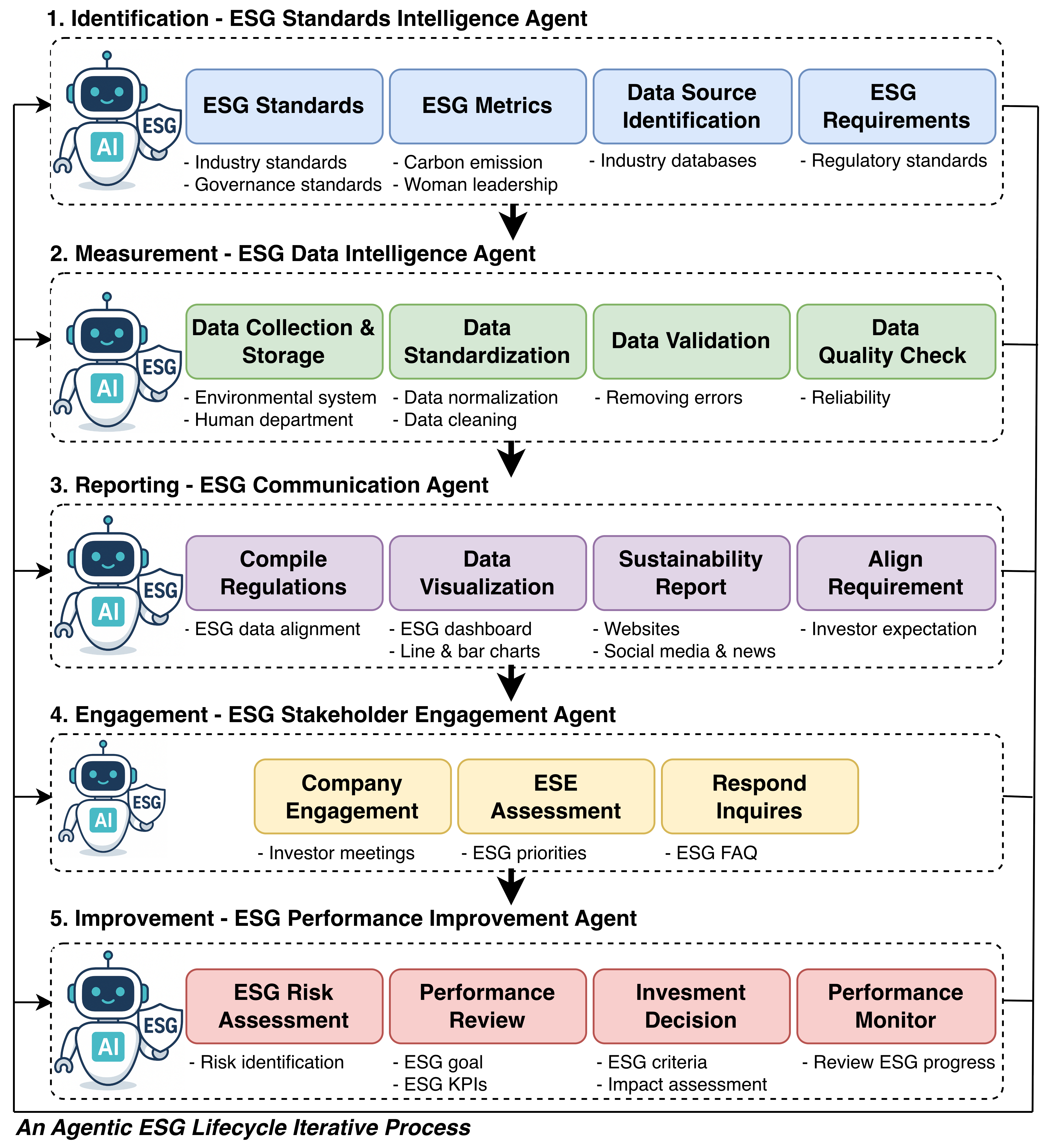}
  \caption{An agentic ESG lifecycle}
  \label{fig:ESG_lifecycle}
\end{figure}

In this section, we present an agentic ESG lifecycle designed to address the key limitations identified in existing ESG lifecycles, i.e., a domain-specific focus, a lack of continuous improvement, and automation capabilities. \add{The core components of our agentic ESG lifecycle are derived based on analysis of ESG reports from organisations, existing ESG lifecycles, as well as major ESG standards, e.g., GRI, SASB, and TCFD. This analysis reveals that organisations must perform regulatory interpretation, conduct data measurement, prepare and disclose ESG reports, engage with stakeholders, and continuously improve their ESG performance to align with these standards. In addition, our proposed agentic ESG lifecycle adopts two main principles observed in advanced agentic AI systems: (1)~agents specialize in specific tasks, and (2)~agents exchange outputs to solve problems and improve performance~\cite{li2023camel, wu2024autogen}.} Figure~\ref{fig:ESG_lifecycle} shows the proposed agentic ESG lifecycle, including five iterative stages: identification, measurement, reporting, engagement, and improvement. We employ a specific AI agent for each stage to address its challenges. These agents interact by exchanging their outputs, thereby improving their performance on ESG tasks across the ESG lifecycle.  Although the lifecycle is iterative, organizations can employ any AI agent they need based on their ESG maturity level and operational priorities. Our proposed ESG lifecycle is an LLM-based agentic ESG system, helping automation and transparency across all stages. The following texts describe each stage in more detail.


\noindent \textbf{3.1. Identification:} The agentic ESG lifecycle begins with the identification stage, where organisations determine the relevant ESG standards, metrics, and data sources applicable to their operations. In this stage, an \textit{ESG Standards Intelligence Agent} (ESIA) is employed to interpret complex ESG standards and identify ESG metrics.

\noindent \textit{3.1.1. Input: The inputs to ESIA include (i) organisational information such as industry sector, market position, or stakeholders' expectations; (ii) ESG standards, e.g., GRI, SASB, and TCFD; and (iii) ESG data collected from industry databases, e.g., the CDP Global Environmental Disclosure Platform\footnote{\url{https://www.cdp.net/en}} and MSCI ESG Ratings.\footnote{\url{https://www.msci.com/our-solutions/esg-investing/esg-ratings}}}

\noindent \textit{3.1.2. Prompt strategy:} ESIA employs a role-based prompting approach to act as an ESG compliance analyst. The prompt integrates organisational information and relevant ESG documents to recommend suitable ESG standards and ESG metrics.

\noindent \textit{3.1.3. Output:} The output of ESIA includes (1) a set of ESG standards aligned with the organisation's profile; (2) a list of relevant ESG metrics grouped under environmental, social, and governance categories; and (3) a list of regulatory requirements.



\noindent \textbf{3.2. Measurement:} This stage focuses on how organizations collect, validate, and analyze their ESG data. In this stage, an \textit{ESG Data Intelligence Agent} (EDIA) assists organizations in extracting ESG information, standardizing data formats, and ensuring data quality across multiple internal sources. 

\textit{3.2.1. Input:} The inputs to EDIA include ESG data from internal sources, such as human resources, financial systems, and operational databases; the selected ESG metrics; and the regulatory checklist generated from the identification stage.

\noindent \textit{3.2.2. Prompt strategy:} EDIA employs a multi-step prompting strategy. First, extraction prompts are used to identify and retrieve ESG data information from multiple internal sources. Second, normalisation prompts guide the model to standardise ESG data according to predefined ESG metrics. Third, validation prompts instruct the model to detect missing, inconsistent, or anomalous values for checking and aligning with the selected ESG standards. Finally, comparative prompts help benchmark against historical ESG datasets to evaluate the quality and completeness of our extracted ESG data.

\noindent \textit{3.2.3. Output:} The output of this stage is a structured and standardized ESG dataset aligned with the ESG metrics and standards. This dataset is used in subsequent stages, i.e., reporting, engagement, and improvement.



\noindent \textbf{3.3. Reporting:} This stage aims to transform the ESG dataset into a comprehensive ESG report for decision-makers and stakeholders. In the agentic ESG lifecycle, an \textit{ESG Communication Agent} (ECA) automates regulatory compilation, visualization, and regulatory alignment, hence minimizing manual effort for disclosing ESG information. 

\noindent \textit{3.3.1. Input:} The inputs to ECA include: (i) the standardized ESG dataset generated in the measurement stage, (ii) the selected ESG standards and compliance checklist defined in the identification stage, and (iii) stakeholder regulatory requirements.

\noindent \textit{3.3.2. Prompt strategy:} ECA employs four prompt strategies. First, we employ regulatory mapping prompts to align ESG data with the selected ESG metrics. Second, visualisation prompts convert ESG data into visual formats such as dashboards, charts, and risk heatmaps, enabling clearer insights into ESG performance and improving the readability for stakeholders. Third, report-generation prompts assist in drafting ESG reports based on the visualisations and performance indicators of the previous step. Finally, alignment prompts ensure the ESG report meets investor expectations, stakeholder requirements, and relevant regulatory obligations.

\noindent \textit{3.3.3. Output:} The outputs of this stage include a generated ESG report containing visual representations of ESG performance indicators, as well as narrative summaries highlighting key trends, risks, and achievements.

\noindent \textbf{4. Engagement:} The engagement stage focuses on strengthening communication and collaboration between organizations and their stakeholders and investors. In this stage, an \textit{ESG Stakeholder Engagement Agent} (ESEA) assists by summarizing discussions, analyzing stakeholder feedback, and responding to ESG-related inquiries. 

\noindent \textit{3.4.1. Input:} The inputs to ESEA include ESG reports from organisations and stakeholders' documents, i.e., meeting transcripts, emails, and responses.

\noindent \textit{3.4.2. Prompt strategy:} ESEA employs summarisation, analysis, and response-generation prompts. The summarisation prompts focus on extracting key expectations and requirements from stakeholders' documents. The analysis prompts aim to align these expectations to specific ESG indicators mentioned in ESG reports. The response-generation prompts provide responses and update frequency for answering-question documents to ensure clarity and responsiveness between organisations and stakeholders.

\noindent \textit{3.4.3. Output:} The outputs of this stage include a summarisation document of stakeholders' requirements and their ESG indicators. 


\noindent \textbf{5. Improvement:} The improvement stage enhances ESG strategies through continuous evaluation, feedback integration, and long-term performance monitoring. In this stage, an ESG Performance Improvement Agent (EPIA) assists by identifying ESG risk, analyzing ESG performance, evaluating investment decisions, and recommending information to keep the ESG report up-to-date. 

\noindent \textit{3.5.1. Input:} EPIA requires (i) an ESG report and ESG data from previous stages, (ii) historical ESG performance data from organisations, (iii) stakeholders' feedback, (iv) predefined ESG risk criteria, and (v) industry benchmarks or investment guidelines. These inputs provide both internal performance indicators and external expectations necessary for performance evaluation.

\noindent \textit{3.5.2. Prompt strategy:} EPIA employs multiple prompts. The risk prompts to detect risk factors in ESG reports and ESG data and may suggest actions to mitigate these identified risks. The evaluation prompts to compare current ESG factors in their reports against historical ESG performance data to identify the ESG factors that need to be improved. The investment prompts to assess how these ESG factors align with investment activities. Finally, the monitoring prompts are employed to track ESG performance over time and raise alerts when potential issues arise.

\noindent \textit{3.5.3. Output:} This stage outputs (i) an ESG risk assessment report with recommended mitigation actions, (ii) a performance evaluation summary highlighting improvement areas, and (iii) ongoing performance monitoring reports. These outputs enable organisations to continuously refine ESG strategies and update ESG reports.

\section{Architecture of LLM-based ESG Framework}
\label{sec:arch}

\subsection{Functional requirements \& Quality attributes}
\label{sec:requirement}

The selection of architectural solutions is guided by a combination of functional requirements and quality attributes~\cite{velasco2024beyond}. Building on the agentic ESG lifecycle, we define four main ESG tasks to support ESG reporting activities. We present functional requirements and describe quality attributes for these tasks. The tasks are key activities in the ESG lifecycle and can be automated, assisted, or fully delegated to modern AI systems. Each task functions as an expert system, which can be implemented using LLMs as the underlying engine. We present them as follows:

\noindent \textbf{1. Report validation and compliance checking:} The tool functions as an expert system that evaluates the compliance of an ESG report with a selected ESG standard. It outputs a completeness score based on the alignment between the report and the standard and provides a list of missing ESG metrics to guide organizations in improving their reports. This task supports the agentic ESG lifecycle by contributing to Stage 1 (identification) and Stage 5 (improvement).  

\noindent \textbf{2. Multi-report comparison:} Given an ESG standard and multiple user-submitted ESG reports, the tool evaluates the completeness and compliance score of each report and provides a ranked list of these reports. This task is related to Stage 4 (engagement) of the agentic ESG lifecycle.

\noindent \textbf{3. Report generation:} The tool generates an ESG report aligned with a given ESG standard and interacts with users to request any required information. This task corresponds to Stage 3 (reporting) of the agentic ESG lifecycle.

\noindent \textbf{4. ESG knowledge base maintenance:} The tool processes ESG documents, i.e., standards, reports, and regulatory requirements, extracts key information, and stores it in a database structured for use in retrieval-augmented generation systems. This task supports Stages 1 (identification) and 2 (measurement) of the agentic ESG lifecycle.

\add{In light of these tasks and the overall agentic ESG lifecycle, we identify eight quality attributes. These attributes were derived from the analysis of ESG reports and standards, and consideration of LLMs' characteristics, i.e., reliability and scalability. The attributes capture the key system requirements necessary to support accurate ESG reporting, effective human interaction, and robust automation across the ESG lifecycle.} \textit{(1)~Accuracy} ensures ESG systems provide correct and reliable outputs. \textit{(2)~Usability} reflects how easily users interact with ESG systems. \textit{(3)~Modularity} enables each ESG task function to operate as an independent component that can be activated based on user decisions. \textit{(4)~Flexibility and adaptability} describe the ability of users to reconfigure ESG tasks, including the integration of AI models to adapt to changing situations. \textit{(5) Scalability} ensures that ESG systems can perform multiple tasks in parallel. \textit{(6)~Fault tolerance} guarantees that the failure of a single AI agent does not compromise the entire ESG system. \textit{(7)~Explainability} ensures that outputs from ESG systems can be explained 
in a human‑understandable manner. \textit{(8)~Answer faithfulness} evaluates how consistent an ESG system's output is with its supporting context or source material. Note that both \textit{explainability} and \textit{answer faithfulness} quality attributes are relevant to LLM-based ESG systems.  

\subsection{Proposed architectures}

\begin{figure}[t!]
    \centering
    \subfloat[]{\includegraphics[height=0.2\linewidth]{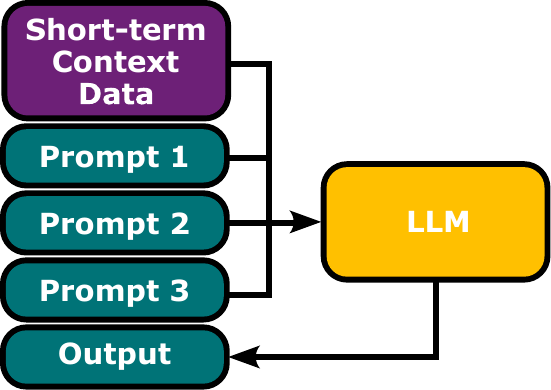}}\\
    \subfloat[]{\includegraphics[height=0.355\linewidth]{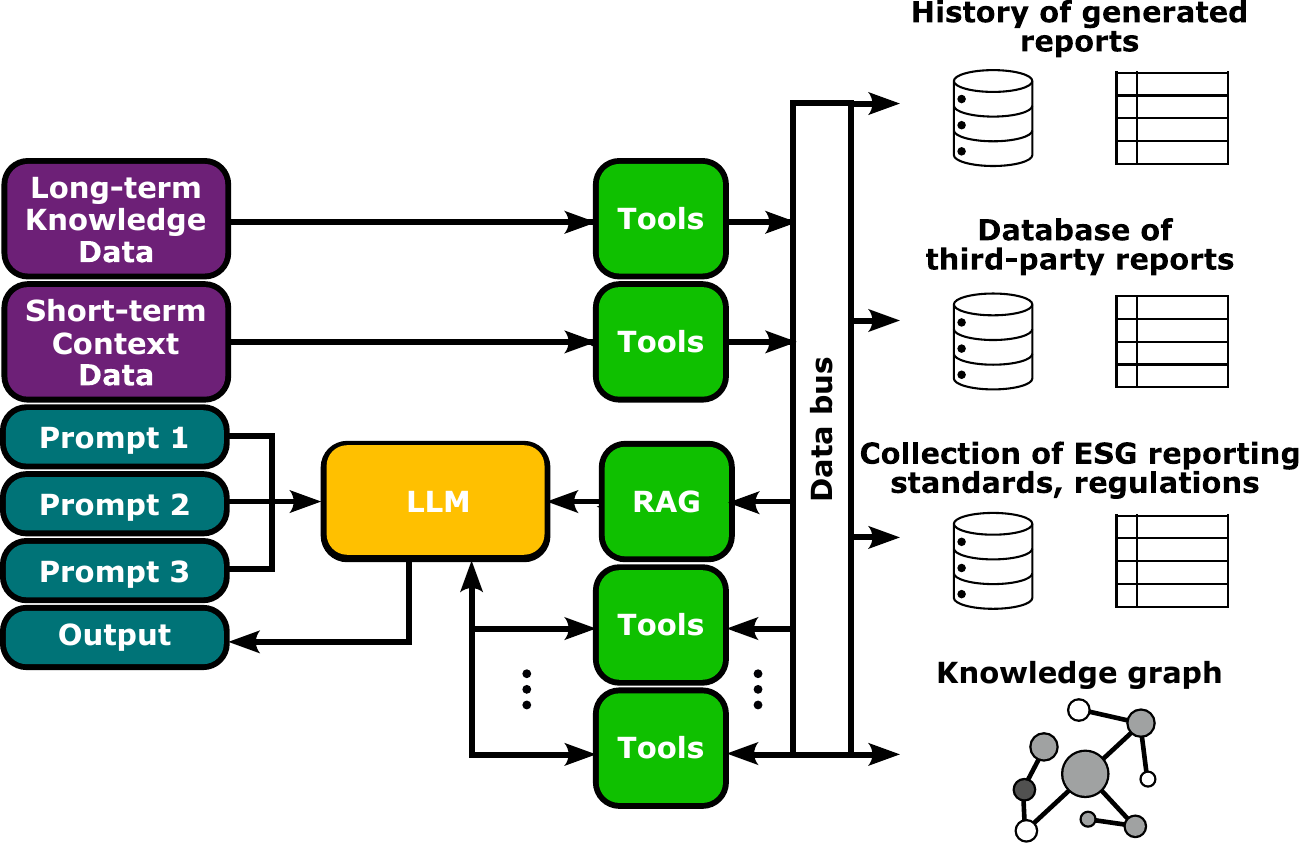}}\\
    \subfloat[]{\includegraphics[height=0.355\linewidth]{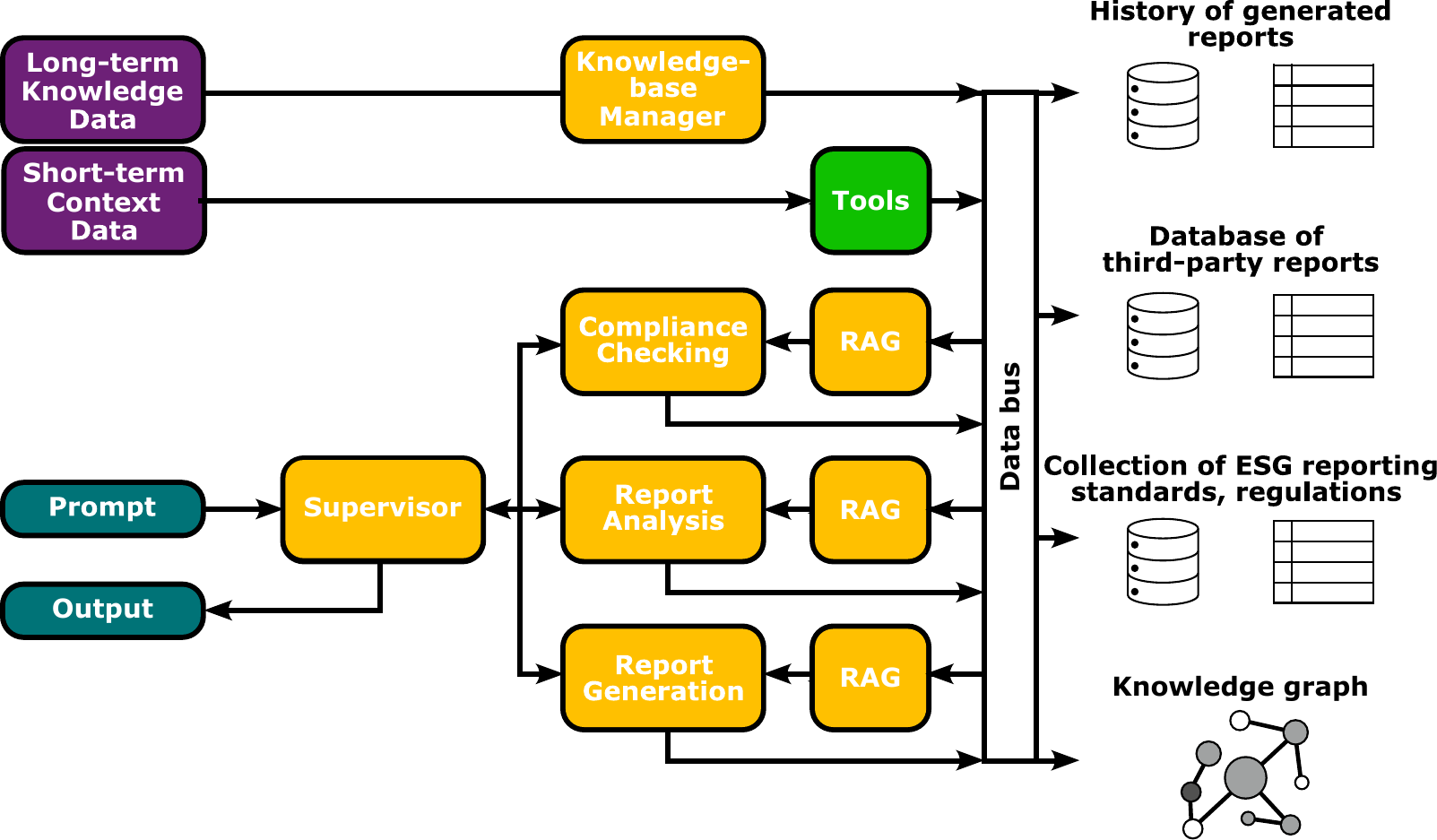}}
    \caption{Architectural approaches for an agentic ESG system: single-model (a), single-agent (b), and multi-agent (c). The colors indicate component roles, with purple, dark blue, yellow, and green representing knowledge sources, prompts, LLM-based agentic components, and non-LLM tools and RAG, respectively.}
    \label{fig:arch}
\end{figure}






\begin{table*}[h!]
\centering
\caption{Performance of three architectural approaches for the ESG report validation task (average results across ESG reports for mean absolute error (MAE), tokens, LLM calls, cost, and energy)}
\begin{tabular}{l|c|c|ccc|c|c|c|c}
\hline
\textbf{Architecture} & \multicolumn{5}{|c|}{\textbf{Mean absolute error across the datasets}}   
 & \multicolumn{4}{|c}{\textbf{Cost metrics evaluated for the Report A}}   \\
\cline{2-10}
\textbf{} & \textbf{\makecell{Report A}} 
 & \textbf{\makecell{Report B}} 
 & \multicolumn{3}{c|}{\textbf{ESG Reports}} 
 & \textbf{\makecell{Tokens}} 
 & \textbf{LLM calls}
 & \textbf{Cost (USD)}
 & \textbf{Energy (kWh)} \\

\cline{2-6}
& \textbf{\makecell{GRI}} 
 & \textbf{\makecell{GRI}} 
 & \textbf{\makecell{GRI}} 
  & \textbf{\makecell{SASB}} 
   & \textbf{\makecell{TCFD}} 
 & 
 &
 & 
 &  \\
 
\hline

Single-model  
    & 51\% & 26\% & 32.30\% & 20.08\% & 23.90\% 
    & 87,541 & 1 
    & \$0.88 & 0.175 \\

Single-agent  
    & 22\% & 6\% & 19.01\% & 21.46\% & 26.95\% 
    & 295 & 2 
    & \$0.03 & 0.006 \\

Multi-agent  
    & 12\% & 4\% & 9.82\% & 16.31\% & 19.59\% 
    & 30,996 & 216 77
    & \$0.03 & 0.061 \\

\hline
\end{tabular}
\label{tab:architecture_evaluation}
\end{table*}

We propose three architectural approaches for addressing the ESG tasks defined in the previous section. Conceptually, each approach corresponds to a specific architectural design: a single-model architecture, handling different tasks using various prompts; a single-agent architecture, combining different prompts with external tools; and a multi-agent architecture, assigning each task to a specific AI agent. 

In this paper, we employ the LangChain framework\footnote{\url{https://www.langchain.com/}} to implement a prototype for each proposed architecture, focusing on the ESG report validation task.\footnote{\url{https://gitlab.com/for_peer_review-group/esg_assistant}} \add{We choose LangChain as it is one of the most popular LLM frameworks, widely adopted by research community and align with our proposed architectures~\cite{apsan2025generating, faiss, wu2024autogen}} For all three architectures, the user interface includes a prompt and a file input, supporting two types of inputs: one for temporary files, i.e., ESG reports, and another for files intended to populate the permanent knowledge base. We explain these architectures in the following paragraphs. 

The single-model architecture employs a single LLM that interacts directly with the user through prompt-based communication (see Fig. \ref{fig:arch}a). It operates without supporting components such as local databases, knowledge graphs, or RAG modules. Consequently, all domain knowledge is embedded solely within the model’s internal weights, and the required context is provided through the prompt. The files provided by the user are converted into string variables and included as part of the prompt. 

In the single-agent architecture (Fig. \ref{fig:arch}b), we enhance an LLM with RAG and tools. File inputs are processed through two pipelines (PDF loaders, parsers, OCR, and text splitters): one for temporary storage and the other for populating the knowledge base. The temporary storage holds uploaded ESG reports and is implemented using vector stores~\cite{faiss}, which enable efficient similarity search for RAG. For the report validation task, this architecture employs a dedicated tool that leverages knowledge encoded as a checklist for each ESG standard, validating the uploaded ESG report against each checklist entry using similarity search. It then assigns ``Yes/No'' labels for each item in the checklist based on a predefined threshold. Finally, the single-agent architecture calculates a compliance score as the ratio of ``Yes'' labels.

The multi-agent architecture (in Fig.~\ref{fig:arch}c) employs agentic RAG and multiple AI agents, each responsible for a specific ESG-related task. Here, we adopt the supervisor architectural pattern,\footnote{\url{https://docs.langchain.com/oss/python/langchain/supervisor}} in which several agents communicate through a central supervisor agent. The supervisor agent interacts with the user, interprets his requests, such as validation, analysis, comparison, generation, or improvement, and orchestrates the appropriate agents. For the validation task, this architecture calls the validation agent to follow the checklist and retrieve relevant information from the document using RAG. Each checklist item is evaluated using an LLM to perform semantic reasoning, resulting in a ``Yes/No'' label. The compliance score is calculated similarly to the single-agent architecture. 

\section{Case study}

In this paper, we apply the proposed architectures to the task of report validation and compliance checking. We create synthetic data, including two ESG reports generated from the GRI standard with LLM assistance, followed by human editing and verification: Report A, which is fully compliant with the GRI standard, and Report B, in which 50\% of the required information was intentionally omitted.
\add{In evaluation, we employ 13 ESG reports and construct ground-truth labels through manual annotation. Specifically, we mapped disclosed ESG factors extracted from each report to the corresponding ESG metrics defined in the GRI, SASB, and TCFD standards.} 
\add{We use mean absolute error, computational cost, and estimated energy as evaluation metrics for the report validation task~\cite{apsan2025generating}. Note that the lower error indicates better performance.} 


\add{Table~\ref{tab:architecture_evaluation} presents the results of the three architectural approaches. The single-model architecture, using GPT-5, shows the highest error across the datasets, with 51\% and 26\% error for Report A and Report B, respectively. It also incurs the highest cost (0.88 USD) and energy consumption (0.175 kWh) due to passing the entire document and a lengthy system prompt. The single-agent architecture, also based on GPT-5, performs well on synthetic reports but less effectively on ESG reports; however, it is the most computationally efficient with the lowest cost (0.03 USD) and energy consumption (0.006 kWh). The multi-agent architecture, employing both GPT-5 and GPT-4o mini, achieves the lowest errors across most datasets, with 12\% and 4\% error for Report A and Report B, and the smallest mean absolute errors for the ESG reports: 9.82\% (GRI), 16.31\% (SASB), and 19.59\% (TCFD); however, its energy consumption is lower than the single-model architecture due to the use of dedicated tools and RAG.} \add{Our paper focuses on architectural design rather than model benchmarking; hence, other LLMs (other than GPT-5 and GPT-4o mini) can be integrated into our proposed architectures. While the results may differ across various LLMs, the tradeoff between single-model, single-agent, and multi-agent remains consistent.}

A key limitation of the single-agent architecture is its development cost. Each tool requires a task-specific heuristic algorithm with parameters that influence overall performance; developers must spend time tuning these parameters, leading to significant manual effort and higher overall cost.
In contrast, the multi-agent architecture delegates most computations to the LLM, minimizing implementation effort. Both these architectures offer strong \textit{modularity} and \textit{flexibility}, allowing new tools or agents to be added easily. In this case study, we notice that the single-model architecture requires more user interactions due to the lack of context and a clear task methodology, leading to increased cognitive load and reduced \textit{usability}.

\section{Related Work}
\label{sec:related_work}



Various ESG standards have been proposed to guide organizations to disclose their ESG performance. Specifically, ESG standards are often grouped into two categories, i.e., industry standards and governance standards. Industry standards, such as GRI \cite{GRI}, SASB \cite{SASB}, and TCFD \cite{TCFD}, define \textbf{\textit{what}} organizations should measure and provide related to three dimensions: environmental, social, and governance. On the other hand, governance standards, i.e., CSRD~\cite{CSRD}, the EU Taxonomy~\cite{EUTaxonomy}, and OECD~\cite{OECDGuidelines}, explain \textbf{\textit{how}} organizations should report and manage their ESG information within legal and policy contexts. 

\add{Although ESG standards provide valuable guidance, they also increase operational complexity. Before employing advanced large language model (LLM) techniques, organizations must first develop a structured understanding of regulatory requirements, define and align concepts across multiple ESG standards, normalize heterogeneous data sources and formats, and design workflows that are compatible with automation. Without this domain knowledge, we limit the effectiveness, reliability, and interpretability of LLM systems. In this paper, we propose an agentic ESG lifecycle that can help domain understanding, harmonization, and workflow formalization, thereby establishing the necessary foundation for integrating LLM capabilities to address ESG problems, i.e., information extraction~\cite{zou2025esgreveal} and text classification~\cite{xia2024using}. We believe that the mentioned challenges, i.e., transforming heterogeneous data formats, structuring complex domain knowledge, and designing automation-compatible workflows, are common across many other domains; hence, the contribution of this study extends beyond ESG, illustrating broader principles for deploying LLM-based systems in real-world organizational settings.}

\section{Conclusions}

In this paper, we propose an agentic ESG lifecycle that bridges the gap between the challenges of generating ESG reports, such as unstructured data, inconsistent terminology, and multi-standard alignment, and the limitations of existing ESG lifecycles, including domain specificity, lack of continuous feedback, and insufficient automation. The proposed lifecycle consists of five stages, i.e., identification, measurement, reporting, engagement, and improvement, in which each stage is supported by an AI agent. Building on this lifecycle, we define four core ESG tasks, i.e., report validation, multi-report comparison, report generation, and knowledge-base maintenance, and present their technical requirements with quality attributes. We then propose three architectural approaches to address these tasks: a single-model, a single-agent, and a multi-agent architecture. These architectures are evaluated using both synthetic and real-world ESG reports. The results show that the single-model architecture is the least accurate and the most costly in terms of computation cost and energy consumption. The single-agent architecture is highly efficient but heavily depends on heuristic tools, while the multi-agent architecture achieves the highest accuracy with moderate computational cost. Future work includes developing an adaptive agent system capable of automatically selecting the most suitable architecture based on ESG task complexity and resource constraints and extending multi-agent capabilities to handle complex ESG tasks. Overall, our paper provides a foundation for integrating AI agents to transform ESG from static reporting activities into dynamic, automated, and accountable sustainability governance. \add{Moreover, the broader contribution of this work lies in demonstrating that deploying LLMs in real-world settings requires deep domain knowledge, the normalization of heterogeneous data sources and formats, and the design of workflows. Without such information, LLMs alone cannot reliably address complex domain tasks.}

\noindent \scriptsize{\textbf{Acknowledgement.} This research/project is supported by the A*STAR under its 2nd CSIRO and A*STAR: Research-Industry (2+2) Partnership Program (Award R24I5IR047). Any opinions, findings and conclusions or recommendations expressed in this material are those of the author(s) and do not reflect the views of the A*STAR.}



%
%
%
\bibliographystyle{splncs04}
\balance
\bibliography{mainref}
\end{document}